\documentclass[12pt,preprint]{aastex}

\shorttitle{Line Polarization within a Giant  \lya\ Nebula}
\shortauthors{Prescott, Smith, Schmidt, \& Dey}

\newcommand{\zLAB}{2.656}
\newcommand{\totsnr}{36}
\newcommand{\Pobs}{3.8}
\newcommand{\Psig}{2.8}
\newcommand{\Pcorr}{2.6}

\newcommand{\lya}{Ly$\alpha$}
\newcommand{\degree}{\ensuremath{^\circ}}
\newcommand{\ergs}{erg~s$^{-1}$}

\begin{document}

\title{The Line Polarization Within a Giant \lya\ Nebula}

\author{Moire K. M. Prescott\altaffilmark{1,2}, Paul S. Smith\altaffilmark{2}, Gary D. Schmidt\altaffilmark{2,3}, and Arjun Dey\altaffilmark{4}} 

\altaffiltext{1}{TABASGO Postdoctoral Fellow; Department of Physics, Broida Hall, Mail Code 9530, University of California, Santa Barbara, CA 93106; mkpresco@physics.ucsb.edu}
\altaffiltext{2}{Steward Observatory, University of Arizona, 933 North Cherry Avenue, Tucson, AZ 85721}
\altaffiltext{3}{National Science Foundation, 4201 Wilson Boulevard, Arlington, VA 22230, USA}
\altaffiltext{4}{National Optical Astronomy Observatory, 950 North Cherry Avenue, Tucson, AZ 85719}

\begin{abstract}

Recent theoretical work has suggested that \lya\ nebulae could be 
substantially polarized in the \lya\ emission line, depending on the 
geometry, kinematics, and powering mechanism at work.  Polarization observations can 
therefore provide a useful constraint on the source of ionization in these systems.  
In this Letter, we present the first \lya\ polarization measurements for a giant \lya\ nebula 
at $z\approx$~\zLAB.  We do not detect any significant linear polarization of the \lya\ 
emission: $P_{Ly\alpha}=\Pcorr\pm\Psig$\% (corrected for statistical bias) within a single large aperture.  
The current data also do not show evidence for the radial polarization gradient predicted by some theoretical models.  
These results rule out singly scattered \lya\ (e.g., from the nearby AGN) and 
may be inconsistent with some models of backscattering in a spherical outflow.  
However, the effects of seeing, diminished signal-to-noise ratio, and angle averaging within 
radial bins make it difficult to put strong constraints on the radial polarization profile.  
The current constraints may be consistent with higher density outflow models, spherically symmetric infall models, 
photoionization by star formation within the nebula or the nearby AGN, resonant scattering, 
or non-spherically symmetric cold accretion (i.e., along filaments).  
Higher signal-to-noise ratio data probing to higher spatial 
resolution will allow us to harness the full diagnostic power of polarization observations 
in distinguishing between theoretical models of giant \lya\ nebulae. 

\end{abstract}


\section{Introduction}
\label{sec:intro}

Very large ($\sim$100~kpc) gaseous nebulae, detectable in the high-redshift universe by 
their spatially extended, luminous \lya\ emission, have received substantial attention as 
the possible sites of ongoing galaxy formation.  The question of what powers these \lya\ 
nebulae (colloquially known as \lya\ `blobs') has been a subject of keen interest since this 
class of objects was first discovered more than a decade ago 
\citep[e.g.,][]{francis96,fynbo99,keel99,stei00}, because the origin of the \lya\ 
presumably holds valuable insight into the dominant physical process at work in these complex 
regions.  Observational studies have uncovered evidence that some \lya\ nebulae are powered by 
gravitational cooling radiation \citep{nil06,smi08}, others by AGN, star formation, or some combination 
\citep[e.g.,][]{basu04,gea09,gea07,mat07,dey05,pres09}.     
Theoretical studies have explored the possibilities of powering by superwind outflows 
and gravitational cooling radiation \citep{tani00,tani01,dijkloeb09,goerdt10,faucher10}.  
However, the uncertainty regarding the degree of obscuration and internal geometry 
coupled with large uncertainties in the modeling of \lya\ radiative transfer has left our 
understanding of the \lya\ nebula phenomenon confused and incomplete.  Additional constraints 
are therefore needed to shed new light on the question of what powers the \lya\ emission.  
In this Letter, we present the first constraints on the \lya\ polarization in a giant \lya\ 
nebula.  

The observable \lya\ polarization from an astrophysical cloud is governed by two important factors 
\citep[see][]{leeahn98,loeb99,ryb99,dijk08a}.  
First, since \lya\ is a resonance line and the primary recombination line of the most abundant element in 
the universe, \lya\ radiative transfer is dominated in most contexts by resonant scattering.  
\lya\ photons are repeatedly absorbed and reemitted by neutral H atoms 
until they are either destroyed via absorption by dust grains or escape the cloud.  
Due to thermal motions of the absorbing atoms, photons execute a random walk 
in both frequency and physical space \citep[e.g.,][]{harr73,neu90,loeb99} and are most able to escape an 
optically thick cloud when they have scattered into the wing of the line profile.  Kinematic offsets 
within the gas - e.g., due to the Hubble expansion, outflows, or infall - can help shift the \lya\ photons 
out of resonance and allow them to escape more readily.   

The second important consideration is the processes that might lead to linear polarization 
of the observed \lya\ emission.  In pure Rayleigh scattering, the phase function is such that a photon 
is twice as likely to scatter forward or backward as it is to scatter at a 90\degree\ angle, 
but the polarization is highest for photons that scatter at 90\degree.  Therefore, 
light that is initially unpolarized can be highly polarized by scattering off neutral atoms if 
scattered through large angles.  In the case of \lya\ scattering, the phase function and degree of polarization 
qualitatively resemble the case of pure Rayleigh scattering, but the details must be calculated quantum 
mechanically \citep[e.g.,][]{brandt59,stenflo80,brasken98}.  A key result is that \lya\ scattering in 
the wing of the line profile yields polarization that is three times higher than \lya\ scattering in 
the line core \citep{stenflo80}.  Thus, those photons that are most likely to escape (those in the line wing) 
are also those with the highest polarization.  If, after the photons have been polarized, however, they 
remain trapped in an optically thick cloud, they will resonantly scatter many more times.  
Since polarization is more likely to be destroyed than preserved in any subsequent scattering event, 
the highest polarization results from single scattering; as a general rule, the {\it more} photons 
scatter, the {\it less} polarization will be observed.  
Robust predictions for more complex scenarios require detailed simulations because the ionization state 
and optical thickness of the cloud, its kinematics, and its geometry will all affect the degree of polarization.  

A number of theoretical papers have modeled the expected level of polarization in the \lya\ line 
under a variety of astrophysical scenarios.  Focusing first on the pre-reionization epoch, \citet{ryb99} 
showed that a cosmological \lya-emitting source embedded within the neutral intergalactic medium (IGM) 
should produce high levels of polarization in the \lya\ emission line and that the polarization should 
increase with radius from the source.  At small radii, where photons are scattered many times before they shift out of 
resonance and escape, the polarization 
is predicted to be low ($\sim14\%$).  Photons that reach large distances from the source, however, 
have typically been scattered in one large jump and are traveling roughly radially.  
The Hubble expansion has shifted them into the wing of the line and their last scattering event has put them 
on a course to the observer, i.e. $\sim$90\degree\ scattering, resulting in higher polarization ($\sim32-60\%$).  

In the post-reionization epoch, the situation is somewhat more complicated, and simulations have thus far 
focused on spherically symmetric geometries.  \citet{dijk08a} showed that in the case of backscattered radiation 
in a spherical galactic outflow, the \lya\ polarization is predicted to increase strongly with radius to levels as 
high as $\sim40\%$, depending on the assumed H\textsc{i} column density.  The high polarization occurs because the kinematic offset of 
the gas shifts the photons out of resonance and into the wing of the line, where the induced polarization is higher and where 
a substantial fraction are able to escape after only one scattering event.  
The radial increase stems from the fact that photons seen from larger radii have scattered through larger angles 
in order to reach the observer.  
In spherically symmetric, collapsing, optically thick clouds, the situation can be similar to 
that of a spherical outflow, with polarization increasing as a function of radius to values as high 
as $\sim35\%$ \citep{dijk08a}.  While photons undergo multiple scatterings in this case, they do so in the line wing 
where polarization is higher.  The radial polarization profile is related to the fact that the \lya\ radiation 
field is increasingly anisotropic with radius; photons emerging from large radii were preferentially traveling 
radially outward prior to scattering, leading to larger scattering angles and higher polarization.  
Finally, in the case of resonant scattering in the IGM, the \lya\ photons scatter a greater number of times and do so in 
the core rather than in the wing of the line, yielding lower polarization \citep[$\sim2-7$\%;][]{dijk08a}.  
Dust was assumed to be minimal in these calculations, but in the presence of significant amounts of dust the polarization 
signature could be higher because the more highly polarized photons are also the ones that scatter the 
least and are therefore more likely to escape before they are destroyed.  
Although these predicted polarization levels are encouraging from an observational perspective, it is 
important to remember that they were derived for the case of spherical symmetry and for specific choices of H\textsc{i} 
column density and velocity profile; in more complicated 
scenarios, e.g., cold accretion along filaments, the polarization may be substantially 
lower \citep{dijkloeb09}.  The \lya\ polarization from these more complex scenarios has yet to be 
rigorously modeled.  

The existing theoretical predictions have thus far lacked an observational response.  In this Letter, 
we present the first constraints on the \lya\ polarization within a giant \lya\ nebula.  In 
Section~\ref{sec:obsredux}, we discuss our observations and data reduction approach.  
Section~\ref{sec:results} presents our results and discusses the implications of our 
finding that the measured polarization of the \lya\ emission in this \lya\ nebula is low 
($\lesssim$5\% overall or $\lesssim$12\% at large radii).  Our conclusions are given 
in Section~\ref{sec:summary}.

\section{Observations \& Reductions}
\label{sec:obsredux}

We chose as our target a giant \lya\ nebula at $z\approx$~\zLAB\ \citep[][hereafter LABd05]{dey05}.  With a diameter of 
roughly 150~kpc and a \lya\ luminosity of $\sim10^{44}$ \ergs, it is one of the 
largest and most luminous \lya\ nebulae known, making it a prime candidate for follow-up observations.  
LABd05 was first discovered thanks to its strong Spitzer/MIPS 24$\mu$m emission, 
stemming from an associated obscured AGN that is offset by $\sim$20~kpc (projected) from the peak of the \lya\ emission 
\citep[][Prescott et al. 2011, in preparation]{dey05}.  

We obtained imaging linear polarimetry centered on the \lya\ line using the Bok 2.3m Telescope and 
the CCD Imaging/Spectropolarimeter, SPOL \citep{schmidt92a}, during the nights of 2007 May 13-15 UT.  
The instrument is a dual-beam polarimeter built around a rotating semi-achromatic half-waveplate and a Wollaston prism.  
The SPOL field-of-view at the Bok Telescope is $100\times100$ pixels = $51\arcsec\times51\arcsec$, 
which for our observations contained both LABd05 and two nearby sources that allowed us 
to verify the position of each dithered frame.

Maximizing the sensitivity of these observations to polarization in the 
\lya\ line required a high throughput narrow-band filter centered at 
\lya\ at the redshift of the nebula.
To achieve this, we selected a filter with an intrinsic central wavelength 
slightly to the red of \lya\ but with high throughput 
($\lambda_{c}=4481.19$\AA, FWHM~$=56.18$\AA, $T_{max}=63.83\%$), 
and used the fact that tilting an interference filter relative to the 
incident beam has the effect of shifting the central wavelength to the blue.  
Using filter transmission measurements taken 
at two different angles, we computed the effective dielectric 
index of refraction of the filter ($n_{l}=2.1$) and used the standard relation 
($\lambda=\lambda_{c}[1-\frac{sin^{2}\theta}{n^{2}_{l}}]^{1/2}$) 
to determine that an inclination angle of $\theta=15\fdg4$ would be needed to shift the 
central wavelength of the filter to match the \lya\ line at this redshift ($\lambda_{Ly\alpha}=4446$\AA).  
We designed an aluminum shim to tilt the filter in the SPOL filter holder 
by this angle during the observations.
Tests done using the instrument as a filtered grating spectrometer at the beginning of the 
run confirmed that the filter bandpass was correctly centered on \lya\ at $z\approx$~\zLAB.  

Over three nights we obtained $18\times3200$s~$=16$~hours of total integration and followed the standard 
procedure for polarimetric observations with dual-band polarimeters.  The observations suffered from 
light to moderate cirrus and poor seeing of 1\farcs5-2\farcs3, which in turn led to increased guiding 
errors, but the advantage of dual-beam polarimetric 
observations is that changes in transmission cancel during the reduction of the data.   
To derive the Q Stokes parameter, we obtained 2 separate Q integrations 
with initial $\lambda/2$ waveplate positions of 0\degree\ and 45\degree, respectively, 
relative to the instrumental zeropoint.  
Each Q integration was divided into 4 individual exposures (200s each), one for each of 4 
equivalent waveplate positions.  After each individual 200s exposure, 
the shutter was closed, the waveplate was rotated by 
$90\degree$, and the shutter was reopened to continue the integration.  
To derive the U Stokes parameter, the same procedure was used to produce 2 separate U 
integrations but with the initial waveplate position set to 
22\fdg5 and 67\fdg5, respectively.  
After each Q and U sequence (3200s), the telescope was dithered by $\sim1-2\arcsec$ in order to 
reduce the effects of flat-field errors on the total flux image.

We reduced the data using IRAF.  We subtracted the bias using 
the overscan region 
but did not apply a dark correction as the dark current was measured to be negligible 
($\sim$0.3 counts per pixel in a 300s exposure).  Domeflats for each Stokes sequence were stacked 
and divided by the median value; the resulting flatfield frame was applied to all images.  
Cosmic rays were removed from individual exposures 
using {\it xzap}.\footnote{Written by Mark Dickinson (NOAO).}  
The standard IRAF-based reduction package for SPOL ({\it impolred}, written by G. Schmidt) 
was applied to the image/waveplate sequences to yield the polarization results.  

We determined the instrumental polarization using observations of unpolarized standard stars (GD319, 
BD+33 2642, and BD+28 4211).  As the instrumental polarization was negligible ($\lesssim0.1\%$), we did not apply 
any correction to the data.  To measure the polarimetric efficiency of the instrument, we observed a Nicol prism 
illuminated with the flatfield lamp through a pinhole aperture.  This effectively 
simulated an observation of a star with 100\%\ polarization and yielded a 
polarimetric efficiency measurement for the $\lambda/2$ waveplate of 97\%\ in this wavelength bandpass.  
The science data were corrected accordingly.   
Observations of interstellar polarization standard stars HD~155528 and Hiltner~960 were 
made to rotate the measured polarization position angle to the equatorial coordinate system \citep{schmidt92b}.  
The dithered offsets between individual frames were calculated using a nearby source within the image 
and used to shift each exposure prior to generating the final image stacks.

\section{Results \& Discussion}
\label{sec:results}

\begin{figure}
\includegraphics[angle=0,width=6.5in]{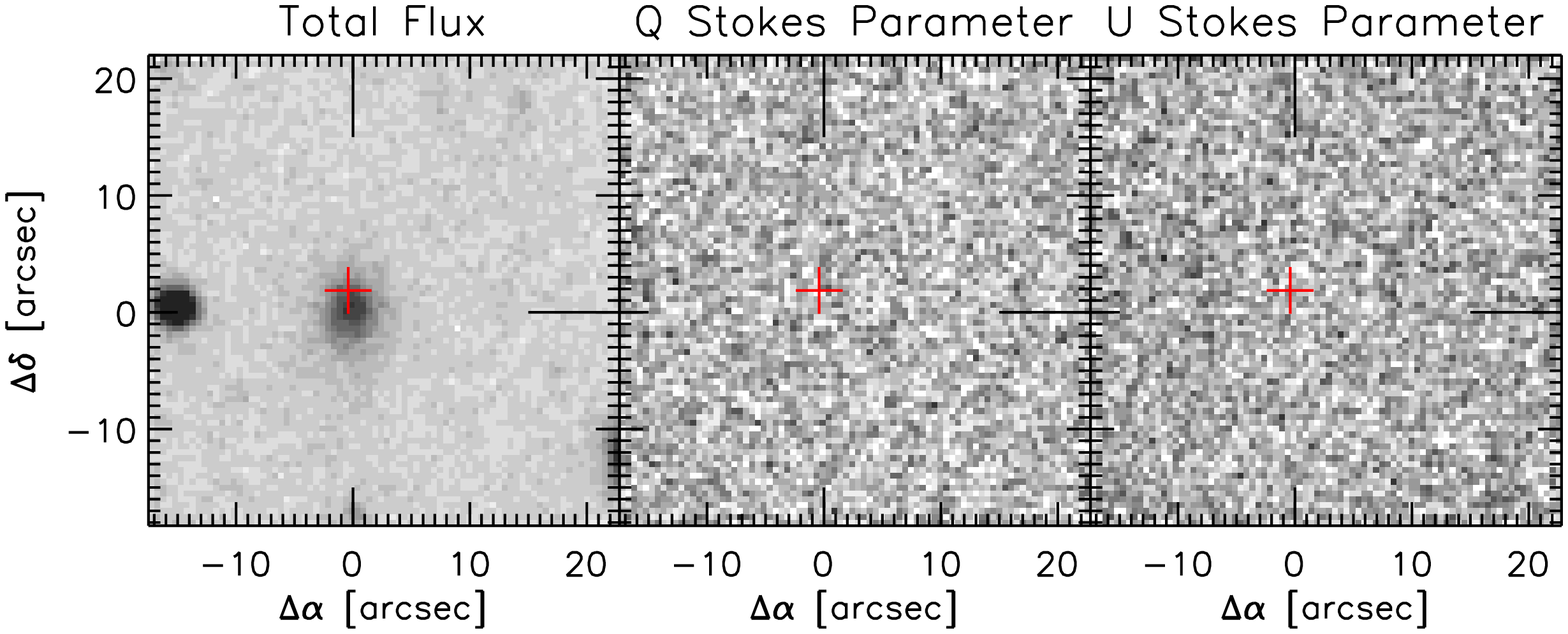}
\caption[Imaging Polarimetry of a \lya\ Nebula.]
{Imaging polarimetry of LABd05.  The position of the \lya\ 
emission peak is indicated by the outer black crosshairs; 
the position of the obscured AGN is shown as a red cross.  
The lefthand panel shows the total flux image, and the 
middle and righthand panels give the Q and U Stokes 
parameter images.  The spatially extended \lya\ emission 
is clearly visible in the total flux image 
($\approx\totsnr\sigma$ in a 8\farcs2 diameter aperture) 
but no polarization is detected (\Pcorr$\pm$\Psig\%).}  
\label{fig:impol}
\end{figure}

The mean total flux image is shown along with the Q and U Stokes parameter images 
in Figure~\ref{fig:impol}.  The \lya\ emission from LABd05 is clearly detected in the total 
flux image at a signal-to-noise ratio of \totsnr\ (16~pix=8\farcs2 diameter aperture).  In the case of 
polarization due to light scattered from the AGN (shown in Figure~\ref{fig:impol} as a red cross 
located at a position angle of 12\fdg5 relative to the \lya\ peak), we would expect the 
electric vector position angle of polarization to be aligned roughly E-W across 
the nebula.  We therefore start by measuring the Q and U Stokes parameters 
within a 16~pix (8\farcs2) diameter circular aperture centered on the peak of the 
\lya\ emission and find that $P_{obs}=\sqrt{Q^2+U^2}=$~\Pobs$\pm$\Psig\%, with the 
uncertainty ($\sigma$) derived from photon statistics.  
As a check, we performed the same aperture photometry on the 18 individual (unstacked) Q and U frames and 
find that the quoted uncertainty yields $\chi^{2}$ indices of unity for the mean Stokes parameters 
derived from these 18 independent measurements.  
When we correct for statistical bias - an important effect 
at low polarization signal-to-noise ratios - the polarization from 
the circular aperture is $P\approx P_{obs}(1-(\sigma_{P_{obs}}/P_{obs})^2)^{1/2}=$~\Pcorr$\pm$\Psig\%, 
where we assume $\sigma_{P_{obs}}\approx\sigma$ \citep{wardle74}.  

Our single aperture measurement of LABd05 is consistent with low \lya\ polarization ($\lesssim$5\%).  
An idealized scenario in which \lya\ photons from the obscured AGN that lies to the north 
are scattering once off a nearby cloud is clearly not allowed by these observations. 

\begin{figure}
\includegraphics[angle=0,width=6.5in]{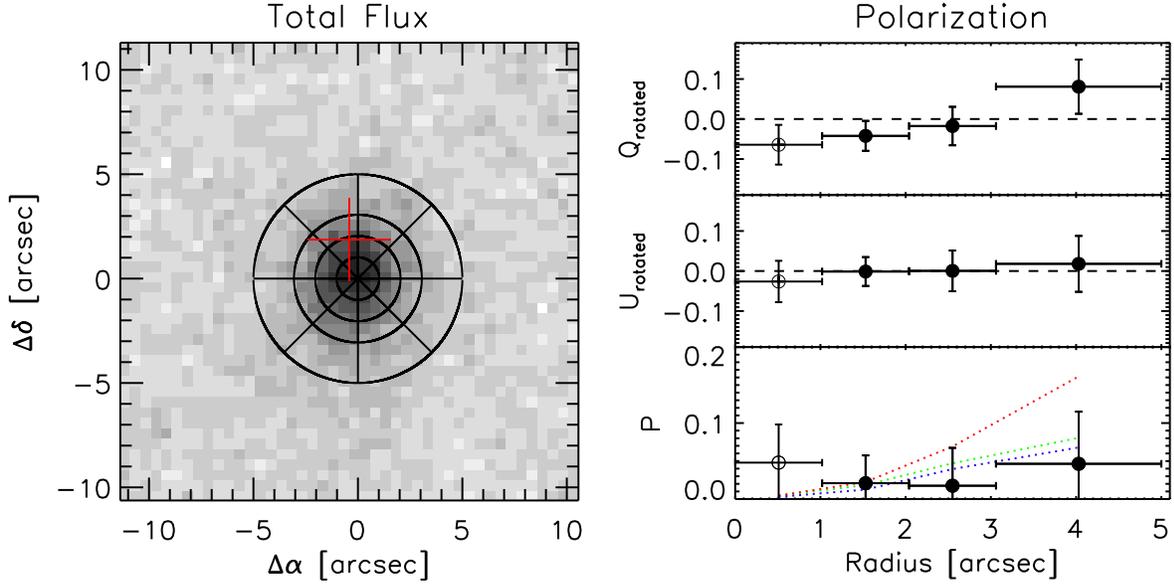}
\caption[Polarization Fraction as a Function of Radius from the Peak of the \lya\ Emission.]
{(Left) Subapertures used to measure the radial polarization profile 
are shown overlaid on the total flux image.  The position of the obscured AGN is shown (red cross) 
at a position angle of 12\fdg5 relative to the center of the \lya\ nebula.  
(Right) Rotated Q and U Stokes parameters and the resulting polarization fraction measured 
within radial bins using a weighted mean and corresponding error.  
Measurements for the innermost radial bin are compromised by poor seeing and 
are denoted with open circles.  Simulated results based on theoretical models are overplotted 
(dotted lines).  Models of backscattering in a spherical outflow - for column densities 
of $N_{H\textsc{i}}=10^{19}$ cm$^{-2}$ (red) and $10^{20}$ cm$^{-2}$ (green) - and 
a spherical infall model (blue) are shown.  Our observations put an upper limit on the observed 
polarization of $\lesssim12\%$ ($1\sigma$) at large radii, corresponding to an {\it intrinsic} 
polarization limit of $\lesssim25\%$.} 
\label{fig:radialpol}
\end{figure}

In the case of a single source of \lya\ photons centrally located within the cloud (e.g., as in the case of spherical 
outflow or infall), the polarization vectors would be oriented perpendicular to the radial vector from 
the center of the nebula (i.e., in concentric rings) and would cancel when measured with a circular aperture covering the entire nebula.  
We therefore measured the Q and U Stokes parameters within subapertures (Figure~\ref{fig:radialpol}) 
and correct the individual subaperture estimates for the position angle of the radial vector to each subaperture.  
In Figure~\ref{fig:radialpol} we plot these rotated Q and U Stokes parameters along with the resulting 
polarization fraction averaged within bins in radius.  Bins were chosen such that each bin has a 
signal-to-noise ratio of $\geq$5 in the total flux image, but we note that the innermost radial bin is 
compromised by poor seeing.  We find that the polarization in each radial bin 
is consistent with zero and that there is no evidence for a radial polarization gradient.  

The lack of significant \lya\ polarization when measured in radial subapertures 
is difficult to interpret due to the effects of seeing, decreased signal-to-noise ratio, 
and angle averaging within individual bins.  
To make a fair comparison, we created simulated observations based on the theoretical models 
of \citet{dijk08a}: two models of backscattering in a spherical outflow with H\textsc{i} column 
densities of $10^{19}$ cm$^{-2}$ and $10^{20}$ cm$^{-2}$, and a model of a spherically symmetric infalling, optically thick 
cloud.  The backscattered outflow model is described in terms of a parameter $\alpha_{c}$, corresponding 
to the outer edge of the outflow.  For the purposes of comparison, we set $\alpha_{c}=5\arcsec$ 
(chosen such that the simulated peak \lya\ surface brightness matches that observed for LABd05, after 
scaling the model appropriately by redshift and luminosity); 
however, our qualitative results are not sensitive to this choice.  
The simulated data were convolved with the typical seeing (FWHM~$=2\farcs0$) and binned radially in the same 
way as the actual data.  The simulated measurements are overplotted in Figure~\ref{fig:radialpol}.  

The observed upper limit on the polarization at the center of the last radial bin is $\leq11.5$\% ($1\sigma$).  
Given the effects of seeing and angle averaging, this corresponds to a $1\sigma$ limit of $\lesssim$25\% 
on the {\it intrinsic} polarization at a radius of $\sim$5\arcsec.  
We find that the observed data are marginally inconsistent with the lower density (higher polarization) 
model of backscattered emission from a spherical outflow, but we cannot rule out higher density outflow scenarios or 
the case of a spherically symmetric infalling cloud or resonant scattering in the local IGM.  

There are two additional possibilities that are likely consistent with low \lya\ polarization.  
The first possibility is \lya\ powered by photoionization from extended star formation or the AGN, which would 
be expected to either be unpolarized (in the optically thin case) or have low polarization similar to the 
resonant scattering scenario.  The second is non-spherically symmetric cold flows: 
\citet{dijkloeb09} point out that \lya\ escaping from one filament is not likely to be polarized 
by scattering off another due to the low predicted volume filling factor of cold filaments.  
However, clear observational constraints on the cold flow phenomenon are not yet available, and rigorous 
modeling of the expected \lya\ polarization in cold flows has yet to be done.

\section{Summary}
\label{sec:summary}

In this Letter, we present the first observational constraints on the \lya\ polarization of 
spatially extended radio-quiet \lya\ nebulae.  The polarization fraction of the \lya\ emission 
within LABd05 is \Pcorr$\pm$\Psig\% measured using a large circular aperture, which is inconsistent with 
a simple model of \lya\ photons (e.g., from the AGN) singly scattering off a nearby cloud.  
We see no evidence for a significant polarization gradient as a function of radius, a result that 
may rule out models predicting high polarization ($\sim40\%$) from backscattering in a spherical galactic outflow.  
However, the effects of seeing, decreased signal-to-noise, and angle averaging within radial bins 
make it difficult to constrain models that result in somewhat lower polarization: higher column density outflow models, 
spherically symmetric infall models, and models that invoke photoionization, resonant scattering, or 
cold flows with non-spherically symmetric geometries.  
These observations provide the first constraint on the \lya\ polarization of giant \lya\ nebulae, but 
as the radial polarization profile is important for distinguishing between models, unleashing the 
full diagnostic power of polarization observations of giant \lya\ nebulae will require higher 
signal-to-noise ratio data and analysis at higher spatial resolution.

\acknowledgments

The authors are very grateful to a number of people for assistance with the imaging polarimetry observations,  
in particular Joseph Scumaci in the Steward Observatory Machine Shop for manufacturing the custom-designed 
filter shim and Heidi Schweiker at NOAO for tracing the narrow-band filter bandpass on short notice, allowing 
us to measure the index of refraction.  We thank Tony Misch, Mike Bolte, Buell Jannuzi, and the 
University of California / Lick Observatory for their assistance in making the narrow-band filter available 
for these polarimetric observations.  The authors are grateful to Kristian Finlator, Mark Dijkstra, and 
an anonymous referee for many helpful discussions and suggestions that improved the paper.  
The authors would also like to thank Mark Dijkstra for sending his theoretical model results.  
M. P. was supported by an NSF Graduate Research Fellowship and a TABASGO 
Prize Postdoctoral Fellowship.  Partial support for M. P. was provided by NASA, through a grant (for program 
GO\#10591) from the Space Telescope Science Institute, which is operated by the Association of Universities for 
Research in Astronomy (AURA) under NASA contract NAS 5-26555.  A. D.'s research is supported by NOAO, which is 
operated by AURA under cooperative agreement with the National Science Foundation. 

\end{document}